\documentclass[12pt,aps,prd,tightenlines,superscriptaddress,preprintnumbers,
   showpacs,nofootinbib]{revtex4}
\usepackage{amsmath}
\usepackage{graphicx}
\usepackage{subfig}
\usepackage{slashed}

\usepackage{color}

\def\neu2{\tilde \chi_2^0}
\def\neu1{\tilde \chi_1^0}

\def\Re{{\cal R \mskip-4mu \lower.1ex \hbox{\it e}\,}}
\def\Im{{\cal I \mskip-5mu \lower.1ex \hbox{\it m}\,}}
\def\ie{{\it i.e.}}
\def\eg{{\it e.g.}}

\def\sub#1{_{\lower.25ex\hbox{$\scriptstyle#1$}}}
\def\tev{\,{\rm TeV}}
\def\gev{\,{\rm GeV}}

\def\dof{\,{\rm dof}}
\def\to{\rightarrow}

\def\subw{_{\rm w}}
\def\mh{\ifmmode m\sbl H \else $m\sbl H$\fi}
\def\mch{\ifmmode m_{H^\pm} \else $m_{H^\pm}$\fi}
\def\ztwo{\ifmmode Z_2\else $Z_2$\fi}
\def\zone{\ifmmode Z_1\else $Z_1$\fi}
\def\mtwo{\ifmmode M_2\else $M_2$\fi}
\def\mone{\ifmmode M_1\else $M_1$\fi}
\def\tb{\ifmmode \tan\beta \else $\tan\beta$\fi}
\def\xw{\ifmmode x\subw\else $x\subw$\fi}
\def\ch{\ifmmode H^\pm \else $H^\pm$\fi}
\def\lum{\ifmmode {\cal L}\else ${\cal L}$\fi}
\def\inpb{\ifmmode {\rm pb}^{-1}\else ${\rm pb}^{-1}$\fi}
\def\infb{\ifmmode {\rm fb}^{-1}\else ${\rm fb}^{-1}$\fi}
\def\epem{\ifmmode e^+e^-\else $e^+e^-$\fi}
\def\ppb{\ifmmode \bar pp\else $\bar pp$\fi}
\def\pbp{\ifmmode ~^(\bar p^)p\else $~^(\bar p^)p$\fi}
\def\bsg{\ifmmode B\to X_s\gamma\else $B\to X_s\gamma$\fi}
\def\bsll{\ifmmode B\to X_s\ell^+\ell^-\else $B\to X_s\ell^+\ell^-$\fi}
\def\bstt{\ifmmode B\to X_s\tau^+\tau^-\else $B\to                                                
  X_s\tau^+\tau^-$\fi}

\newskip\zatskip \zatskip=0pt plus0pt minus0pt
\def\matth{\mathsurround=0pt}
\def\lsim{\mathrel{\mathpalette\atversim<}}

\def\atversim#1#2{\lower0.7ex\vbox{\baselineskip\zatskip\lineskip\zatskip
  \lineskiplimit
  0pt\ialign{$\matth#1\hfil##\hfil$\crcr#2\crcr\sim\crcr}}}

\def\sigv{\ifmmode \langle\sigma v\rangle\else $\langle\sigma                                     
  v\rangle$\fi}
\def\tsigv{\ifmmode \langle\sigma v\rangle R^{2}\else $\langle\sigma                              
  v\rangle                                                                                        
  R^{2}$\fi}
\def\Dxx{\ifmmode D_{xx} \else $D_{xx}$\fi}
\def\Dpp{\ifmmode D_{pp} \else $D_{pp}$\fi}
\def\ddp{\ifmmode \frac{\partial}{\partial p} \else
  $\frac{\partial}{\partial p}$\fi}
\def\alle{\ifmmode (e^{+}+e^{-}) \else $(e^{+}+e^{-})$ \fi}
\def\pamr{\ifmmode e^{+}/(e^{+}+e^{-}) \else $e^{+}/(e^{+}+e^{-})$
  \fi}
\def\pbarp{\ifmmode \bar{p}/p \else $\bar{p}/p$ \fi}

\def\boverc{\ifmmode B/C \else $B/C$ \fi}
\def\chisq{\ifmmode \chi^2 \else $\chi^2$ \fi}
\def\rchisq{ \ifmmode \chi^2/\dof \else $\chi^2/\mathrm{dof}$ \fi}

\def\sir{ \ifmmode \sigma_{SI,p}\!\times\! R \else $\sigma_{SI,p}\!\times\! R$ \fi}
\def\sdr{ \ifmmode \sigma_{SD,p}\! \times \! R \else $\sigma_{SD,p}\! \times\! R$ \fi}
\def\Olsp{ \ifmmode  \Omega h^2|_{\rm LSP} \else $$ \Omega h^2|_{\rm LSP} \fi}
\def\Owmap{ \ifmmode  \Omega h^2|_{\rm WMAP} \else $\Omega h^2|_{\rm WMAP}$ \fi}
\def\capr{ \ifmmode  C_c \else $C_c$ \fi}

\def\uns{ \ifmmode \Upsilon (nS) \else $\Upsilon (nS)$ \fi}
\def\ups{ \ifmmode \Upsilon \else $\Upsilon$ \fi}

\begin{document}

\thispagestyle{empty}

\preprint{UCI-HEP-TR-2013-10}

\title{Particle Physics Implications and Constraints on Dark Matter Interpretations of the CDMS Signal
\vspace*{0.5cm}}

\author{Randel C. Cotta}
\affiliation{Department of Physics and Astronomy,
University of California, Irvine, CA 92697, USA \vspace*{0.5cm}}

\author{Arvind Rajaraman}
\affiliation{Department of Physics and Astronomy,
University of California, Irvine, CA 92697, USA \vspace*{0.5cm}}

\author{Tim M.P. Tait}
\affiliation{Department of Physics and Astronomy,
University of California, Irvine, CA 92697, USA \vspace*{0.5cm}}

\author{Alexander M. Wijangco}
\affiliation{Department of Physics and Astronomy,
University of California, Irvine, CA 92697, USA \vspace*{0.5cm}}

\date{\today}

\pacs{95.35.+d, 14.70.Bh}

\begin{abstract}
Recently the CDMS collaboration has reported an excess of events in the signal region of a search for dark matter
scattering with Silicon nuclei.  Three events on an expected background of 0.4 have a significance of about $2\sigma$,
and it is premature to conclude that this is a signal of dark matter.  Nonetheless, it is important to examine the
space of particle theories capable of explaining this excess, to see what theories are capable of explaining it, and how one
might exclude it or find corroborating evidence in other channels.  We examine a simplified model containing a scalar
mediator particle, and find regions consistent with the CDMS observations.  Bounds from colliders put important
restrictions on the theory, but viable points, including points leading to the observed thermal relic density, survive.
\end{abstract}

\maketitle
\newpage

\section{Introduction}

Astronomical and cosmological probes of dark matter not only exist, but indicate that dark matter is five 
times as prevalent in the universe than the conventional forms of matter described by the Standard Model \cite{Ade:2013zuv}. 
Despite this abundance however, knowledge of dark matter remains perplexingly incomplete. Principle among 
these unknowns are the mass of the dark matter (DM) particle and the nature of its interactions with the 
Standard Model (SM), both of which are unconstrained over many orders of magnitude.   

A diversity of theoretical models has grown to accompany the diversity of allowed phenomenology \cite{Feng:2010gw}. Extremely light and 
weakly-coupled axions \cite{Kim:2008hd,Rosenberg:2000wb} are a canonical scenario of DM with phenomenology that differs drastically from that of
the more usually discussed WIMPs \cite{Bertone:2004pz,Jungman:1995df,Chung:2003fi,Servant:2002aq,Cheng:2002ej,Hooper:2007qk},
 though essentially arbitrary phenomenology can be obtained from hidden sector
models \cite{Pospelov:2007mp,Pospelov:2008jd,Feng:2008ya,Feng:2008mu}, which may be designed to solve problems unrelated
to dark matter ($\eg$, generation of cosmic baryon number \cite{Kaplan:2009ag,Falkowski:2011xh,Davoudiasl:2012uw}).

Given this diversity, the experimental effort to measure such interactions has become increasingly creative. In addition to the
traditional three-pronged experimental program consisting of direct detection, which seeks to measure DM-nucleon scattering, 
colliders searches for DM production and indirect detection searches for the energetic products of DM annihilation in
astroparticle experiments, are studies of even more diverse effects, $\eg$, observed and simulated shapes of DM halos \cite{Lin:2011gj}, the
detailed nature of the CMB \cite{Galli:2011rz,Hutsi:2011vx,Finkbeiner:2011dx} and primordial element abundances \cite{Cyburt:2004yc} and cooling of astrophysical objects \cite{Raffelt:1999tx}.

Recently, the CDMS collaboration has made the interesting observation of an excess of 3 events over an expected background of
$0.4$ events, that can be interpreted as a signal detection with $\sim 2\sigma$ significance. Such a result is clearly inconclusive 
on its own and should be subjected to the utmost scrutiny, especially as the favored mass $m_{DM}\simeq 8.6\gev$ coincides with 
the sensitivity threshold of the experiment. Despite these considerations, the result is very interesting in light of 
similar anomalous results, such as from CoGeNT \cite{Aalseth:2010vx}, and in the favored-region's proximity to the predictions of some 
well-motivated theoretical models \cite{Fitzpatrick:2010em}.

Describing a light DM particle with such (relatively) large interactions with the SM and that wouldn't have already been seen 
elsewhere is a phenomenological challenge. There exist several ``portals'' (in effective operator language: SM-singlet operators
 built only out of SM fields) by which such DM may easily communicate with the SM, each of which may naturally suggest vector,
scalar or fermionic mediators and have been studied in some detail in the context of light DM 
\cite{Hooper:2012cw,Andreas:2010dz,Fitzpatrick:2010em,Falkowski:2009yz,Okada:2013cba,Choi:2013fva}. In this work we will consider a
generic model of Dirac fermionic DM interacting with the Standard Model
via a relatively light scalar mediator particle. For such a model to avoid being ruled out from the outset we consider our mediator
to be coupled to SM fermions in minimal-flavor-violating (MFV \cite{D'Ambrosio:2002ex}) fashion, suggesting a natural connection
 between the physics that
generates our DM and messenger to the physics of the Higgs sector and electroweak symmetry breaking. We will describe regions
of parameter space for which our model obtains scattering in the range of the CDMS result, where the annihilations in our model
are sufficient for equalling the cosmological DM relic density and regions that are already ruled out by collider and low-energy
experiments.

The rest of this paper is divided into four sections. In Section \ref{model} we describe and discuss our simplified
model framework, in Section \ref{dmpheno} we describe our model's DM phenomenology, in Section \ref{colliders}
we describe collider and low-energy bounds that can be placed on the parameter space of our model and in Section \ref{discuss}
we present a concluding discussion.

\section{Simplified Model Framework}
\label{model}
We work in the framework of a simplified model consisting of the Standard Model supplemented by
a Dirac DM particle $\chi$ and a CP-even scalar messenger $\phi$.  Since the CDMS signal is suggestive of
a WIMP whose mass is well below $M_Z / 2$, we restrict ourselves to considering dark matter which is an electroweak
singlet in order to avoid large contributions to the invisible width of the $Z$ boson \cite{Beringer:1900zz}.  
Fitting the CDMS signal region will imply ${\mathcal O}(0.1 - 1)$ coupling between $\phi$ and $\bar{\chi}\chi$,
suggesting that $\phi$ should also be an electroweak singlet.  
The mass of the $\chi$ is fixed by the CDMS
signal to $m_\chi \simeq 8.5$~GeV.  In the discussion below, we fix the dark matter mass to this value and comment
where appropriate as to how our results would change for different masses.

In order to evade very strong bounds from
flavor-violating observables, we invoke minimal flavor violation \cite{D'Ambrosio:2002ex}
with regard to the $\phi$ coupling to quarks,
\begin{eqnarray}
\mathcal{L}_{int} = g_{\chi} \phi \bar{\chi} \chi + \sum_i g_d \lambda^d_i \phi \bar{d}_i d_i
+ \sum_i g_u \lambda^u_i \phi \bar{u}_i u_i
\end{eqnarray}
where $\lambda^d_i$ and $\lambda^u_i$ are the down-type and up-type Yukawa interactions.  In addition to the
masses $m_\chi$ and $m_\phi$, the model is specified by the dimensionless
couplings to dark matter $g_\chi$, to down-type quarks (scaled by the appropriate Yukawa interaction) $g_d$, and
similarly defined coupling to up-type quarks $g_u$.
In what follows we will work primarily in the 3-dimensional space ($m_{\phi}$, $g_{\chi}$, $g_d$).  We consider
two distinct cases for $g_u$:
\begin{itemize}
\item $g_u\sim 1.8\; g_d$, leading to iso-spin preserving (IP) elastic scattering in direct detection experiments; or
\item $g_u\sim -1.015\; g_d$, leading to isospin-violating (IV) scattering with $f_n/f_p\sim -0.7$, designed
to maximally weaken the sensitivity of Xenon-based searches \cite{Feng:2011vu}.
\end{itemize}
It is worth noting that even for $g_u \sim \; g_d$, the elastic scattering cross section will be similar
for protons and neutrons, owing to the relatively small contribution of the up and down quarks because of
their small Yukawa interactions.  One could also write down (and put bounds on)
a coupling between $\phi$ and leptons, but such
an interaction is largely orthogonal to a discussion of the CDMS signal.  Where relevant, we 
will comment on the bounds on such a coupling below.

There are also potentially renormalizable interactions between $\phi$ and the Standard Model Higgs doublet, $H$.  
In general, the details of the scalar potential are not very important for the phenomena of interest here, and we leave
a detailed analysis for future work.  However, it is worth noting that mixing between $\phi$ and the Higgs
boson allows for $\phi$ to be produced via typical Higgs production modes, including $\phi Z$ at LEP II.
For masses less than about 110 GeV, null results of Higgs searches at LEP generically imply
that the mixing is no larger than ${\mathcal O}(10\%)$ \cite{Searches:2001ad}, although there are windows of mass
where bounds are weaker, and might even be interpreted as not very significant hints for a
positive signal \cite{Dermisek:2007yt}.

While we remain agnostic as to the origin of the simplified model framework, it is worth noting that one can imagine a
simple UV-completion of the scalar sector
based on a two Higgs doublet model augmented by a gauge singlet scalar.  The two Higgs
doublets provide sufficient freedom in the Yukawa couplings so as to realize $g_u$ and $g_d$ in the desired ranges,
with the (mostly singlet) $\phi$ inheriting the couplings through modest mixing with a combination of the physical
CP even Higgs bosons.   
Perhaps the most studied model containing these ingredients is the NMSSM \cite{Ellwanger:2009dp,Balazs:2007pf}.
It has been pointed out that one can find limits in the NMSSM parameter space that attain large scattering cross-sections
with a low DM mass \cite{Draper:2010ew,Carena:2011jy,Cao:2011re} although there may be some tension with other constraints
as, in supersymmetric models like these,
large cross-sections tend to come hand-in-hand with sizable couplings to $W^{\pm}/Z^0$ \cite{Das:2010ww}. Variations of supersymmetric
models consisting of the MSSM plus a singlet super-field can realize suitable cross sections 
\cite{Hooper:2009gm,Belikov:2010yi,Buckley:2010ve}. For an example of a non-supersymmetric UV completion see \cite{He:2013suk}.

\section{DM Observables}
\label{dmpheno}
In this section we focus on finding regions of our parameter space that are attractive from a DM standpoint: light DM with large 
elastic scattering cross-sections. Although we are particularly interested in scattering, we also calculate relic density and
 discuss current annihilation cross-sections for our $\chi$ to give a sense of the cosmological history necessary in such a scenario. We 
consider messenger masses in a wide range, $1\gev\lsim m_{\phi} \lsim 100\gev$, anticipating (as is confirmed below) that mediator masses
 above $\sim 100\gev$
will be non-trivially constrained by collider monojet searches\footnote{For mediator masses heavier than typical LHC center-of-mass
energies the limit should be essentially the same as the stringent EFT bounds derived in \cite{Rajaraman:2011wf}}. We use \textit{MicrOMEGAs v2.4} \cite{Belanger:2010gh}
for all elastic scattering and annihilation cross section calculations.


For our direct detection calculation we use a local DM density $\rho_0=0.3\gev / \mathrm{cm}^3$ and nuclear form factors:
\begin{eqnarray*}
f^p_u=0.023,\hspace{10 mm} f^p_d&=&0.033,\hspace{10 mm} f^p_s=0.05,\\
f^n_u=0.018,\hspace{10 mm} f^n_d&=&0.042,\hspace{10 mm} f^n_s=0.05.
 \label{formfactors}
\end{eqnarray*}
Appropriate values for the strange-flavored scalar form factors are hotly-debated at current 
\cite{Ellis:2008hf,Gasser:1990ap,Bernard:1993nj,Pavan:2001wz,Young:2009zb,Thomas:2011cg,Junnarkar:2013ac,Alarcon:2012nr}, the choice 
$f^N_s\approx 0.05$ is on the low side of proposed values, making it a conservative choice for our purposes. The 
uncertainty coming from the strange quark is anyway not critical for our purposes: we consider a wide range of elastic
scattering cross-sections\footnote{This range corresponds to the lower-most and upper-most values on the $2\sigma$
ellipse of the result \cite{Agnese:2013rvf}.}, 
\begin{eqnarray}
10^{-6} \mathrm{pb}\lsim\sigma_{\mathrm{SI,N}}\lsim 3*10^{-4} \mathrm{pb},
\label{ddrange}
\end{eqnarray}
as interesting for our purposes. The scattering cross section depends on the couplings only through the product, $g_{\chi}g_d$.

We calculate the thermal relic density of our $\chi$ assuming that the only relevant processes at freezeout are those in 
our simplified model. As always, this is a fairly heavy-handed assumption and may or may not be relevant in any particular
completion of our model. Despite this, our thermal relic calculation remains useful for denoting regions of parameter space where extra
theoretical structure\footnote{$\eg$, non-thermal evolution or dark sector states that participate in the thermal calculation} may be 
necessary to increase or decrease the relic density with respect to our minimal scenario and where our model saturates the 
\textit{Planck} collaboration's measurement \cite{Ade:2013zuv}, $\Omega_{CDM}h^2\approx 0.1146$, on its own. Annihilations proceed through t-channel
 $\chi\bar{\chi}\to \phi\phi$ (when kinematically
 available) and through s-channel $\chi\bar{\chi}\to f\bar{f}$, the former depending on the couplings only as $g_{\chi}^2$
 and the latter as $g_{\chi}g_d$. Both of these processes are actually p-wave processes at leading order (suppressed by
 $\upsilon^2\sim10^{-6}$) so current annihilations from our simplified model are predicted to be much below the
 canonical $\sigv\sim 3*10^{-26} \mathrm{cm}^3/\mathrm{s}$. Similarly low rates are calculated in the resonant region 
 $2m_{\chi}\approx m_{\phi}$, although \textit{Planck}-level relic densities are achieved for much lower coupling values. 
If our model were to also include a pseudo-scalar state, $a$, then there would be available s-wave processes giving current
 annihilations close to the canonical value\footnote{As may be desired given the current (inconclusive, but interesting)
hints of $\sim 10\gev$ DM particles annihilating to $b$'s or $\tau$'s conributing to the $\gamma$-ray spectrum at the 
Galactic Center \cite{Abazajian:2012pn,Hooper:2011ti}.}. Such pseudo-scalars are not hard to come by theoretically 
($\eg$, in approximately SUSY-preserving multiplets) and would have no effect on scattering rates (momentum suppressed)
 but potentially sizable effects on the other observables, such as collider production.

\begin{figure}[hbtp]
    \centering
    \includegraphics[width=1.0\textwidth]{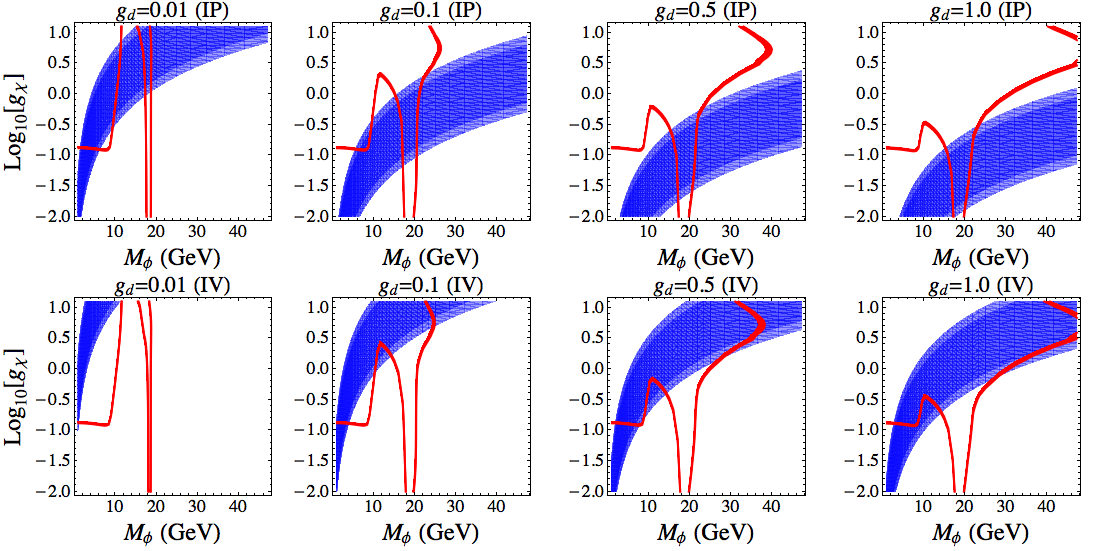}
    \caption{Spin-Independent Scattering and Relic Density. The blue band denotes SI scattering cross-sections within the range Eqn. \ref{ddrange} 
      (darker and lighter regions describing the extent of $1\sigma$ and $2\sigma$ ellipses in the result \cite{Agnese:2013rvf}, respectively). The red band shows 
    where our $\chi$'s relic density is $\Omega_{\chi}\approx\Omega_{CDM}$. In the upper panels $g_u$ and $g_d$ are related such that $f_n=f_p$ (IP), 
    while in the lower panels $f_n/f_p=-0.7$ (IV).}
    \label{ddrd}
  \end{figure}

In Figure \ref{ddrd} we map out the combinations of $g_{\chi}$ and $m_{\phi}$ for which scattering cross sections are within the 
range Eqn.\ \ref{ddrange} and for which the relic density matches the \textit{Planck} value for both IP and IV cases and for several values of $g_d$.
The features of the relic density band are easy to understand: there is a sharp upturn where the $\chi\chi\to\phi\phi$ channel becomes
phase space suppressed ($m_{\phi}\approx m_{\chi}$) and a sharp downturn in the resonant annihilation region ($m_{\phi}\approx 2m_{\chi}$).
Annihilation cross-sections (not shown) are $\sigv\lsim 3*10^{-30} \mathrm{cm}^3/\mathrm{s}$ on the \textit{Planck} band.
In the IV case, scattering cross-sections are reduced by destructive interference and we observe a shift of the favored region
for scattering toward larger coupling values. We observe regions where both large scattering cross-sections and $\Omega_{\chi}\approx\Omega_{CDM}$
can be obtained simultaneously, for essentially any choice of $g_d$. While this happens both for very light mediators ($m_{\phi}< 10\gev$) and
for very heavy mediators ($m_{\phi}> 20\gev$), we expect these regions to be in danger either from $\ups$-decay data or from collider searches. In contrast,
regions of overlap in the $m_{\ups} < m_{\phi} < 2m_{\chi}$ range are particularly hard to constrain.

\section{Collider \& Low-Energy Constraints}
\label{colliders}
\subsection{Mono-Objects}

Intuition garnered from DM effective theory analyses over the last few years suggests that collider searches may have the 
final say on the viability of this scenario \cite{Beltran:2010ww,Goodman:2010yf,Goodman:2010ku,Bai:2010hh,Rajaraman:2011wf,Fox:2011fx,Fox:2011pm,Bai:2012xg,Cotta:2012nj,Carpenter:2012rg,Bell:2012rg}.
 Such searches typically look for DM direct production by studying single objects
(monojets, monophotons, etc.) recoiling off of a missing transverse momentum vector and, unlike direct detection experiments, remain 
sensitive to arbitrarily small DM masses. The caveat to these searches is the 
efficacy of the EFT description, which can give either an overly-conservative or an overly-optimistic sense of the
collider reach in light-mediator scenarios. For our mediators, with the DM mass fixed at $m_{\chi}=8.5\gev$, there are roughly
 three regimes for collider production: \textit{(i)} the mediator is very heavy compared to typical machine center-of-mass energies, 
\textit{(ii)} the mediator is light compared to collider center of mass energies but heavier than $2m_{\chi}$ and \textit{(iii)}
the mediator is lighter than $2m_{\chi}$. Scenario \textit{(i)} is the regime where the EFTs should give basically the right answer,
in scenario \textit{(ii)} the mediator can be produced on-shell so we would expect the EFT bounds to be conservative relative to
the exact bounds and in scenario \textit{(iii)} the mediator can never be put on-shell, the production cross-section is a rapidly
falling function of the mono-object's $p_T$ and the EFT bounds would suggest much tighter constraints than what one would actually
get in the full calculation. Of course these regimes bleed into each other a bit, here we seek to describe this behavior.
For studies involving light vector mediators, see Refs.~\cite{An:2012va,Frandsen:2012rk,An:2012ue,Shoemaker:2011vi}.

Here we focus on LHC monojet searches, which we expect to provide the tightest constraints in this class of experiments. Monojet
bounds from the Tevatron were checked ($c.f.$, \cite{Bai:2010hh}) as well and they are not competitive with those coming from the LHC\footnote{Monophoton 
bounds from LEP are irrelevant unless our mediator were to have large couplings to the electron, which seems unlikely in our
construction.}. We mimic cuts from the ATLAS analysis \cite{ATLAS:2012ky} and use the typical \textit{MadGraph}(v5)-\textit{Pythia}(v6)-\textit{PGS}(v4) 
chain \cite{Alwall:2011uj,Sjostrand:2006za,pgs} (hereafter \textit{MPP}) with default ATLAS detector card to simulate signal and background rates.
   \begin{figure}[hbtp]
    \centering
    \includegraphics[width=0.8\textwidth]{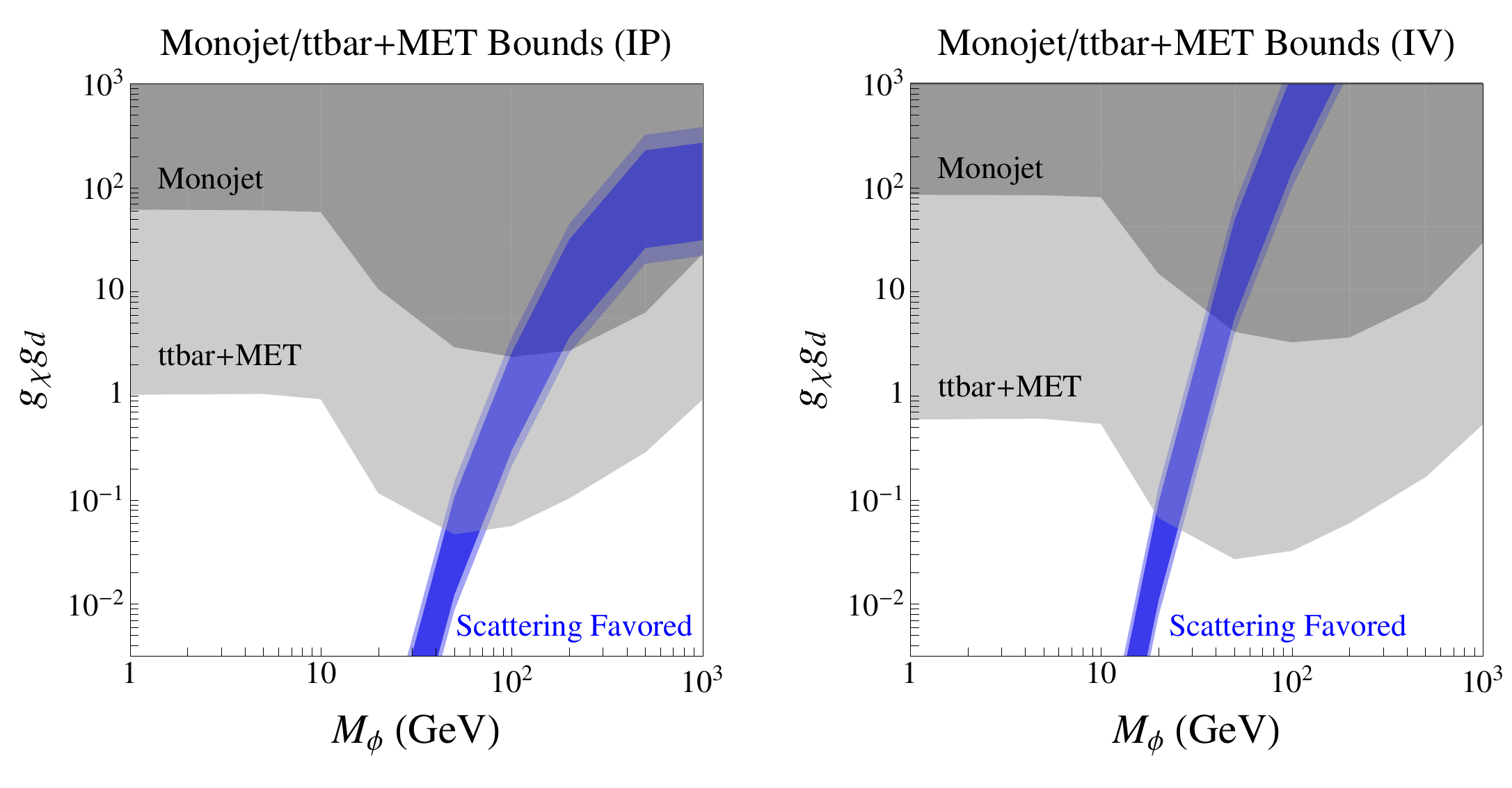}
    \caption{
      Monojet and $t\bar{t}+\mathrm{MET}$ bounds on our model in the $g_{\chi}g_d$ vs.\ $m_{\phi}$ plane (IP left panel, IV right panel).
      The blue bands gives scattering cross-sections in the desired
      range, as in Fig.\ \ref{ddrd}, while the gray regions are excluded by the ATLAS monojet search \cite{ATLAS:2012ky} and the ATLAS
      $t\bar{t}+\mathrm{MET}$ search \cite{ATLAS-CONF-2012-167} (both at 95\% confidence) as noted in the figure. The limits in this plot
      were generated with fixed $g_d=1$.}
    \label{1jet}
  \end{figure}
Monojet bounds are presented in Figure \ref{1jet}. The features of these curves can be easily understood: The cross-section is highly suppressed and nearly constant
in the $m_{\phi}<2m_{\chi}$ regime where the mediator cannot be put on-shell. The kink occurs at $m_{\phi}=2m_{\chi}$ whereafter the monojet bounds become more and
more constraining until the eventual fall off above typical center-of-mass energies. We know that our couplings must increase with the mediator mass in order
to have scattering cross-sections in the range Eqn.\ \ref{ddrange}, here we see that our model will actually run into monojet constraints before reaching its
ultimate perturbativity bound at $g_{\chi}g_d\sim 4\pi$. Interestingly however, Figure \ref{1jet} shows that the monojet reach is much less than that from
the heavy-flavor $t\bar{t}+\mathrm{MET}$ search for all $m_{\phi}$, this is what we will describe next.

\subsection{Heavy-Flavor Searches}

While the MFV structure of our messenger's couplings keep direct collider production of $\phi$'s highly-suppressed, the
large couplings to top and bottom quarks suggest large rates for $\phi$'s radiated off of the final states in heavy flavor (HF) production. Since our
$\phi$'s may be made to decay either dominantly to missing transverse energy (for $g_{\chi}\gg g_{d}$) or to $b\bar{b}$ (for $g_{d}\gg g_{\chi}$),
 heavy flavor searches both 
with and without associated MET may be applicable. HF searches with MET are typical of the suite of SUSY searches for 
third-generation squarks ($\eg$, \cite{ATLAS-CONF-2012-167}), while HF searches without MET are not nearly as common.
An example of the latter is the search for signals of Higgs production in the $t\bar{t}H\to t\bar{t} b\bar{b}$ channel 
(in practice, the $t\bar{t}+$b-jet channel \cite{ATLAS-CONF-2012-135,Aad:2013tua}). Here we investigate bounds on our model's
 parameter space that can be derived from these two searches. Another recent work that considered heavy-flavored final states 
and dark matter is \cite{Lin:2013sca}

The ATLAS analysis \cite{ATLAS-CONF-2012-167}, uses $13~\mathrm{fb}^{-1}$ of $8\tev$ data to place very stringent
 constraints, $\mathcal{O}(1 \mathrm{fb})$, on $t\bar{t}+\mathrm{MET}$ from BSM sources. Here we use the full 
\textit{MPP} analysis chain to simulate the SM background to this 
search and to get a sense of the acceptance profile for tagging the two tops in our signal. To calculate the
 signal rate we assume that the acceptance (more precisely, the part of which comes from top-tagging) for signal
 events is essentially the same as that for the SM background. This allows us to do an initial calculation of the 
signal at parton level, before applying the more involved $m_{T2}$ cut to accurately reproduce the MET acceptance 
(the quantity that is really sensitive to the kinematics of our signal events) in reasonable computational time. The
particular MET and $p_T$ cuts that we used were those of the ``110 SR'' signal region defined in \cite{ATLAS-CONF-2012-167}. 
The resulting excluded region is described in Figure \ref{1jet} and is seen to be stronger for all $m_{\phi}$ than that from
 the monojet search. Our model's mediator mass is bounded to be $m_{\phi}\lsim 45\gev$ (IP) or $m_{\phi}\lsim 20\gev$ (IV), 
in both cases far smaller than the model's ultimate perturbativity bound $g_{\chi}g_d\lsim 4\pi$.

\begin{figure}[hbtp]
    \centering
    \includegraphics[width=0.8\textwidth]{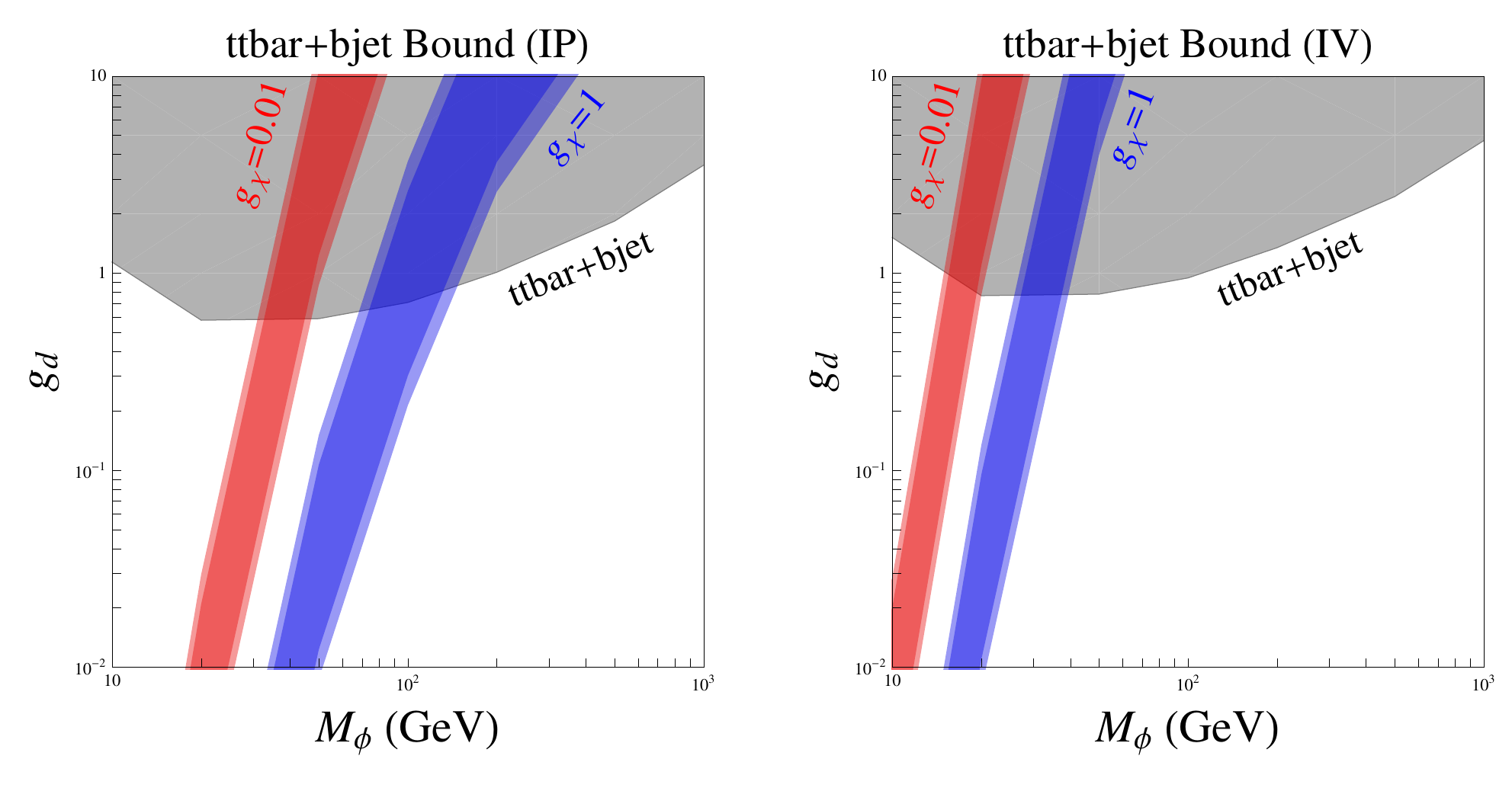}
    \caption{Heavy Flavor bounds on our model in the $t\bar{t}b$ channel (IP left panel, IV right panel). As this search depends only on 
      the coupling $g_d$ we display, in red and blue bands, the favored regions for scattering with $g_{\chi}=0.01$ and $g_{\chi}=1$, respectively.
      The gray region denotes parameter space for which the $t\bar{t}\phi$ production cross-section is greater than that for $t\bar{t}H$ production
      of the SM Higgs (our rough criterion for exclusion given the result \cite{ATLAS-CONF-2012-135}).}
    \label{hfvis}
  \end{figure}

In the $t\bar{t}b$ channel it is more difficult to obtain an accurate bound in our parameter space. The most relevant\footnote{
The analysis \cite{ATLAS-CONF-2012-135} uses a similar data sample but is too focused on the SM Higgs to be useful in bounding our model.}
analysis in this regard is the ATLAS measurement \cite{Aad:2013tua} of the ratio of $t\bar{t}b$ and $t\bar{t}j$ (where $b$ denotes a b-tagged
jet and $j$ denotes all jets) in $4.7~\mathrm{fb}^{-1}$ of $7~\tev$ data. The result is not easy to interpret as a bound in the present context,
as the measured ratio $t\bar{t}b$/$t\bar{t}j$ is found to be in excess of the SM expectation at the $1.4\sigma$ level. Rather than
trying to interpret this as evidence for new physics, we simply suppose that the measurement is roughly consistent with the SM (including a 
$125\gev$ Higgs) prediction and require that our model not contribute to $t\bar{t}b$ at a level greater than that from the Higgs. We 
calculate both the $t\bar{t}\phi$ and $t\bar{t}H$ cross-sections using \textit{MPP} with the ``nominal''
sample selection cuts described in \cite{Aad:2013tua} to determine the ``excluded'' regions for which the $t\bar{t}\phi$ cross-section is greater
than the $t\bar{t}H$ cross-section. The result is described in Figure \ref{hfvis}, where the excluded region is compared to the preferred
regions for scattering with two choices for $g_{\chi}$, $g_{\chi}=0.01$ and $g_{\chi}=1$ (as the $t\bar{t}\phi$ cross-section depends only on $g_d$).

\subsection{B-Factory Constraints}
\begin{figure}[hbtp]
    \centering
    \includegraphics[width=0.8\textwidth]{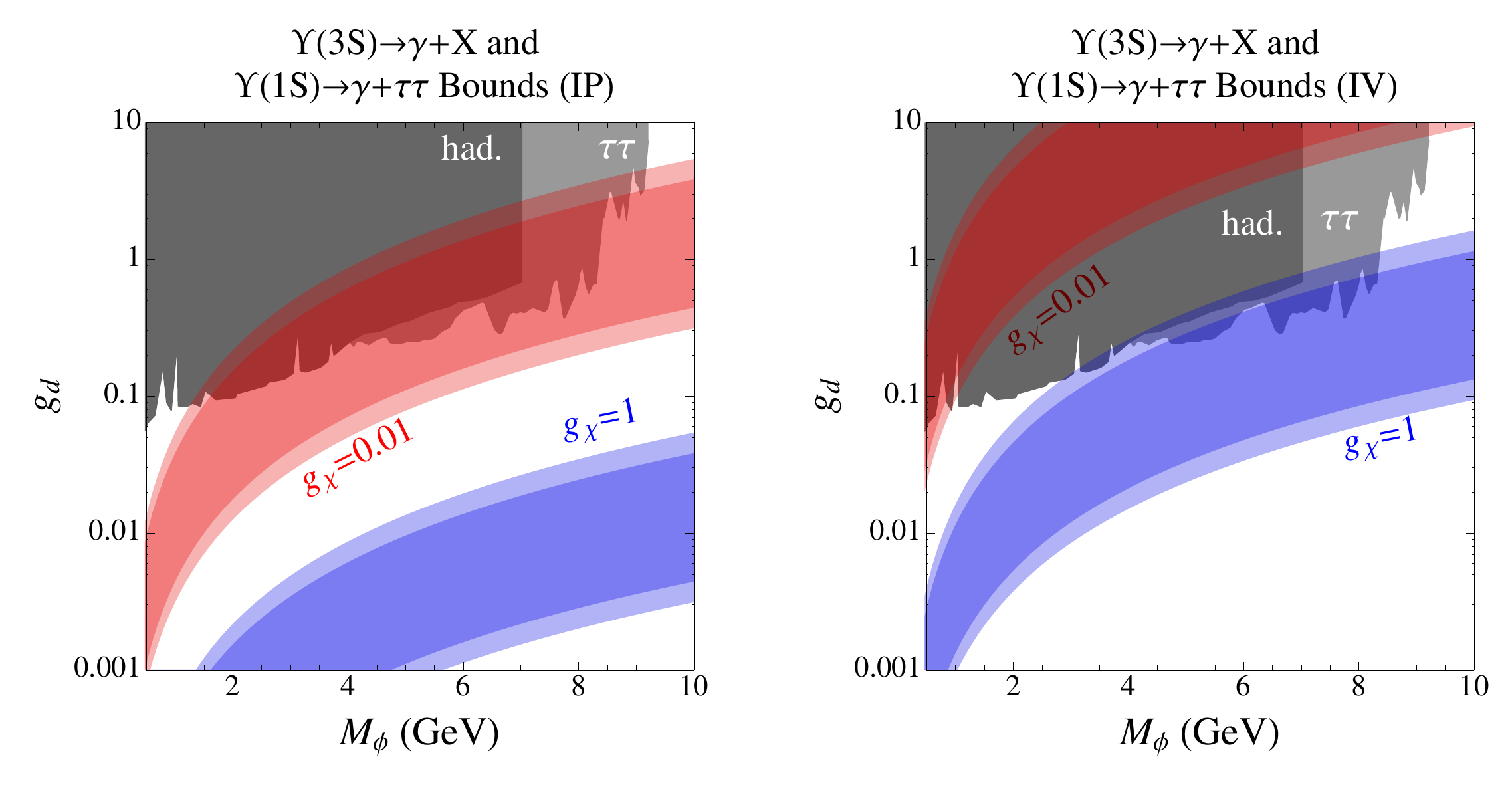}
    \caption{
      Bounds from the radiative $\ups$-decays to hadronic and di-tau final states. Gray regions are excluded by \emph{BaBar}
      analyses \cite{Lees:2011wb} and \cite{Lees:2012te} (as noted in the figure) at 90\% confidence. The red and blue bands
      give direct detection favored regions for $g_{\chi}=0.01$ and $g_{\chi}=1$ as in Fig.\ \ref{hfvis}. Favored regions are
      calculated for both IP (left) and IV (right) cases.}
    \label{ups}
  \end{figure}

For mediators with $m_{\phi}\lsim m_{\Upsilon}\approx 10\gev$ one must consider the possible signatures of our model in $\uns$ 
decay processes. Since our DM has $2m_{\chi}>m_{\Upsilon}$ we do not expect signatures in $\ups$ decays with invisible products (although
these would become relevant for $m_{\chi}\lsim 5 \gev$), instead we consider radiative $\ups$ decays, $\uns\to\gamma\phi\to\gamma X$ 
where\footnote{Of course, ``$\phi$'' here refers to our mediator, not the light unflavored meson.} $X$ is some visible system recoiling
off of a monochromatic $\gamma$. We consider two \emph{BaBar} collaboration analyses: \cite{Lees:2011wb}, a search for photon resonances in
 $\Upsilon (3S)\to \gamma + \mathrm{hadrons}$ and \cite{Lees:2012te}, a search for photon resonances in $\Upsilon (1S)\to \gamma + \tau^+\tau^-$.
Both of these results provide a bound on $g_d$ (independent of $g_{\chi}$), the former considering only quark coupling while the latter requires the 
model-dependent assumption that $g_l=g_d$. We calculate the associated rates in our model space, following closely the work \cite{Yeghiyan:2009xc}.
The resulting bounds are shown in Figure \ref{ups}. The $\ups$ data limits the
$g_d$ coupling to be generally $g_d\lsim 0.1$ for models with $m_{\phi}\lsim 10~\gev$, ruling out favored parts of parameter space where
$g_{\chi}$ is small. There is a large dependence on the choice of IP or IV scattering, the latter being constrained much more tightly at a
 given scattering cross-section by the $\Upsilon$ data. 

\subsection{Exotic Higgs Decays}

Given the necessarily small mixing between our messenger and the SM Higgs, we expect that the 
current constraint on the Higgs invisible width (about $40\%$, per \cite{Bai:2011wz}) is not tight enough to constrain our model. If 
our mediator is light, $m_{\phi}\ll m_{H}$, then, as in many NMSSM discussions, we may imagine producing a
pair of boosted $\phi$'s and searching for pairs of boosted objects from their decays.
While the rate of such events depends on the details of the UV physics that give rise to our simplified model, the resultant
striking signature may be the first place in which such a model can be discovered.

An example of such an analysis is the ``ditau-jet'' search strategy, outlined in \cite{Englert:2011iz}, wherein one tries to discern
``jets'' composed of a pair of boosted $\tau$'s ($\eg$, coming from the $\phi$ decays) from generic QCD jets. In this work it was 
demonstrated that (with consideration of a jet's $p_T/m_j$ ratio and application of jet-substructure techniques) one can tag ditau-jets
with high-efficiency and low-mistag rates. It was argued that, for a light-scalar model with nearly exactly the same kinematics as ours, 
an appropriate series of cuts would yield effective signal and background cross-sections of $\sigma_s=0.5~\mathrm{fb}$ and $\sigma_b=0.12~\mathrm{fb}$,
and thus a $S/\sqrt{B}=5$ discovery for $\mathcal{L}=12~\infb$ of $14\tev$ LHC data. In our model, if we assume that down-type quarks \emph{and}
 down-type leptons are both normalized with the parameter $g_d$, then $BR(\phi\to\tau\tau)\sim\mathcal{O}(10\%)$. Given this then, even assuming a
 scalar trilinear coupling $g_{h\phi\phi}=\sqrt{4\pi}$, our model would be far from detectable in such a search. If, however, the lepton couplings
are normalized independently of $g_d$ then, with $g_l$ such that $BR(\phi\to\tau\tau)\sim\mathcal{O}(100\%)$ , our model would also be observable in
$\mathcal{L}=12~\infb$ of $14\tev$ LHC data.

\section{Discussion}
\label{discuss}
We have investigated diverse bounds on the parameter space of a simplified model of DM whose phenomenology could plausibly explain the
low-mass and high-cross-section signal of DM scattering in the CDMS Silicon data. Our model is typical of some extensions of the SM Higgs
sector that give light scalars coupling to SM fermions in an MFV pattern ($\eg$, coupling like a Higgs). We have shown that such models
can easily attain the necessary large scattering cross-sections for couplings of $\mathcal{O}(0.1-1)$, while also attaining the correct relic
density, in many regions of this subspace. If such a model were to be supplemented with a pseudoscalar of similar mass to our messenger $\phi$,
essentially none of the above story would change qualitatively, except that one would have the kind of canonical s-wave annihilation rates
that we may already be seeing in the Galactic Center.

\begin{figure}[hbtp]
    \centering
    \includegraphics[width=0.9\textwidth]{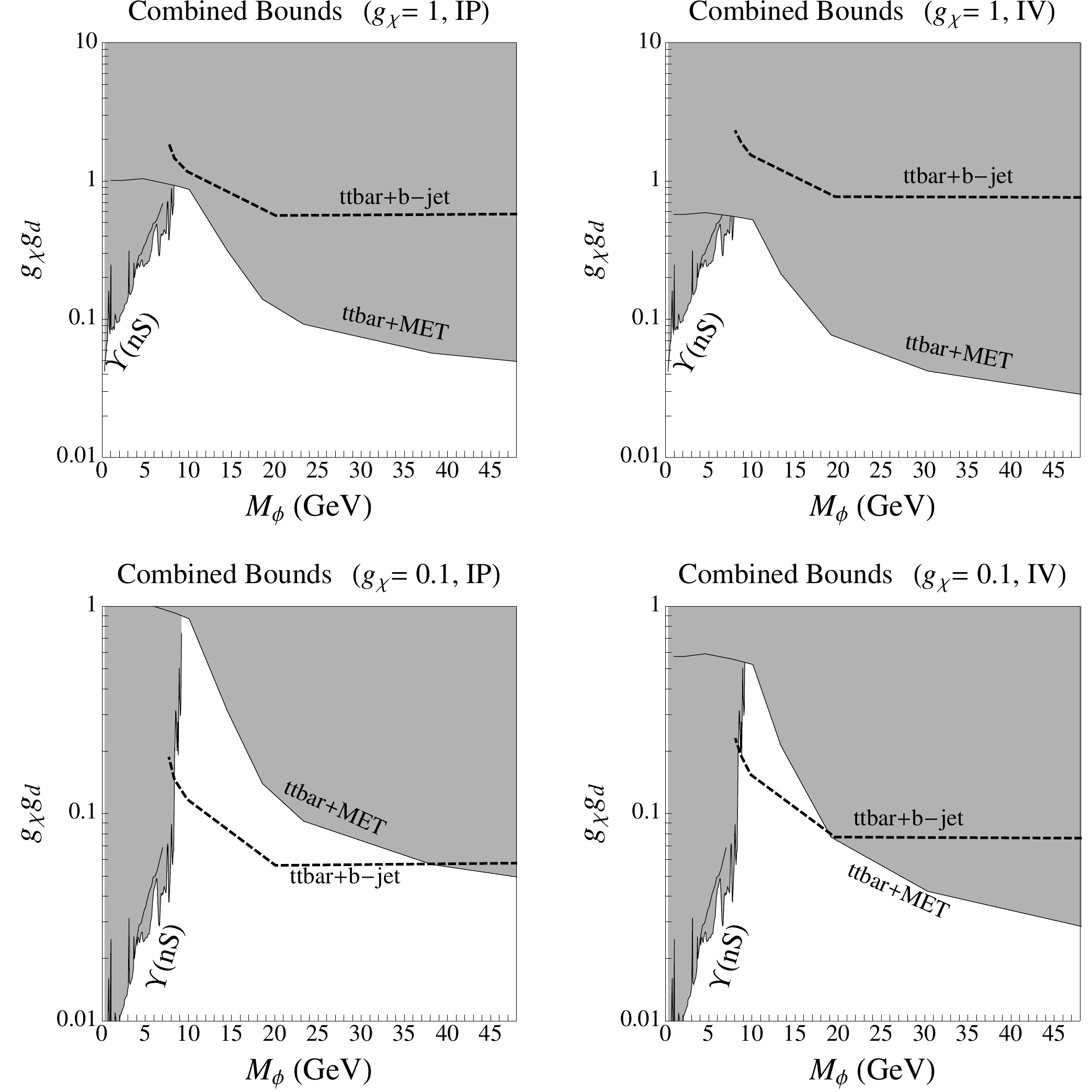}
    \caption{Combined bounds in the $g_{\chi}g_d$ vs.\ $m_{\phi}$ plane. Bounds from $t\bar{t}+\mathrm{MET}$, $t\bar{t}+$b-jet and radiative $\Upsilon$ decays
      (in both hadronic and $\tau$ channels) are labelled accordingly. Monojet bounds are irrelevant, given the axes ranges plotted. We choose $g_{\chi}=1$
      ($g_{\chi}=0.1$) in the upper (lower) panels to translate bounds that only depend on $g_d$ onto this plane. Left and right panels correspond
      to IP and IV scenarios, respectively. We use a dashed line to remind the reader that the $t\bar{t}+$b-jet bound is particularly rough (as described
      in the text).}
    \label{combo1}
  \end{figure}

\begin{figure}[hbtp]
    \centering
    \includegraphics[width=0.9\textwidth]{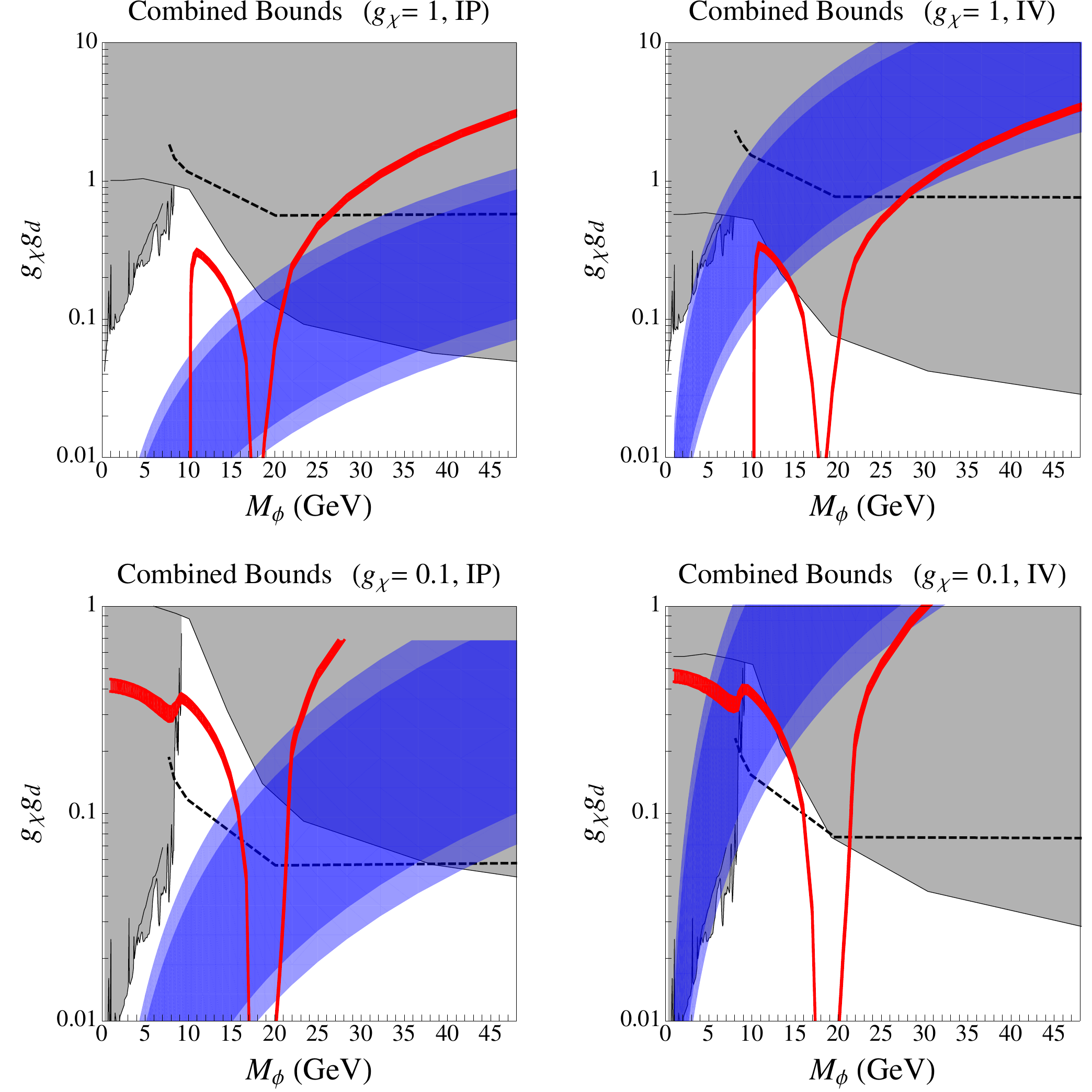}
    \caption{As in Figure \ref{combo1}, but with the inclusion of direct detection and \textit{Planck} favored bands in blue and red, respectively.}
    \label{combo2}
  \end{figure}

We have discussed collider and low-energy B-factory bounds on our parameter space and the complementarity of these bounds. A round-up of these
results is described in Figures \ref{combo1}-\ref{combo2}, where all bounds are collected and plotted in the $g_{\chi}g_{d}$ vs.\ $m_{\phi}$ plane.
Results are given for two different choices of $g_{\chi}=1$ and $g_{\chi}=0.1$. In Fig.\ \ref{combo1} we find that, for large $g_{\chi}=1$, the
combination of $t\bar{t}+\rm{MET}$ and $\Upsilon(nS)$ data require $g_d\lsim 0.1$ except in the difficult region
$m_{\Upsilon(3S)}<m_{\phi}<2m_{\chi}$ where $g_d\lsim 1$. For smaller $g_{\chi}=0.1$ we see that the $t\bar{t}+$b-jet bound (depending only on $g_d$)
supplants the $t\bar{t}+\rm{MET}$ bound (depending on $g_{\chi}g_{d}$) to require $g_d\lsim 1$ for all $m_{\phi}$. In Figure \ref{combo2} we overlay
the favored regions for scattering and relic density in our parameter space. We see that the isospin-violating case is more highly constrained
than the isospin-preserving case, owing to the generally larger product $g_{\chi}g_{d}$ required to produce scattering signals at the CDMS level.

The fact that a light DM particle and scalar messenger coupling \textbf{so strongly} to SM fermions is even phenomenologically viable at this point
is very interesting. It is completely plausible that a model like ours could be discovered first in 
direct detection experiments (as it may already have been!), especially
for mediator masses in the difficult range $m_{\Upsilon(3S)}<m_{\phi}<2m_{\chi}$. From what we have shown it is also plausible that such a discovery could be
corroborated (or such a model ruled out) by LHC searches for anomalous heavy flavor final states, strongly motivating a more careful look
at such signatures under more generic ($\ie$, than SM Higgs or MSSM sparticle) expectations.

\newpage
\section*{Acknowledgments}
The authors would like to acknowledge helpful discussions with J.~Shelton, J.~Zupan, C. Wagner, and L. Tao. The research of R.C.C. and A.R. is
supported by the National Science Foundation under grant PHY-0970173. The research of T.M.P.T. is supported in part by NSF
grant PHY-0970171 and by the University of California, Irvine through a Chancellor's fellowship.

\newpage
\bibliographystyle{JHEP}
\bibliography{bib}

\end{document}